\documentclass{article}%
\usepackage{amsfonts}
\usepackage{amsmath}
\usepackage{amssymb}
\usepackage{cite}
\usepackage{charter}
\usepackage{graphicx}%
\providecommand{\U}[1]{\protect \rule{.1in}{.1in}}

\begin{document}

\title{Higher-order properties and Bell-inequality violation for the three-mode
enhanced squeezed state}
\author{Shuang-Xi Zhang$^{1}$\thanks{shuangxi@mail.ustc.edu.cn }, Hong-Chun Yuan$^{2}%
$\thanks{\emph{Corresponding author}, yuanhch@sjtu.edu.cn}, Hong-Yi
Fan$^{1,2}$\\$^{1}${\small Department of Material Science and Engineering, }\\{\small University of Science and Technology of China,\ Hefei, Anhui 230026,
China}\\$^{2}${\small Department of Physics, Shanghai Jiao Tong University, }\\{\small Shanghai, 200030, China}}
\maketitle

\begin{abstract}
By extending the usual two-mode squeezing operator $S_{2}=\exp \left[
i\lambda \left(  Q_{1}P_{2}+Q_{2}P_{1}\right)  \right]  $ to the three-mode
squeezing operator $S_{3}=\exp \left \{  i\lambda \left[  Q_{1}\left(
P_{2}+P_{3}\right)  +Q_{2}\left(  P_{1}+P_{3}\right)  +Q_{3}\left(
P_{1}+P_{2}\right)  \right]  \right \}  $, we obtain the corresponding
three-mode squeezed coherent state. The state's higher-order properties, such
as higher-order squeezing and higher-order sub-Possonian photon statistics,
are investigated. It is found that the new squeezed state not only can be
squeezed to all even orders but also exhibits squeezing enhancement comparing
with the usual cases. In addition, we examine the violation of Bell-inequality
for the three-mode squeezed states by using the formalism of Wigner representation.

\end{abstract}



\section{Introduction}

Widespread attention has been paid in recent years to a class of optical field
states that are called squeezed states \cite{1}, where the fluctuation in one
quadrature is less than that of the vacuum state \cite{2}. Squeezed states
have manifestly nonclassical properties, which may find application in
low-noise optical communications as well as in the gravitational-wave
detection \cite{3}. The two-mode squeezed state \cite{3-0}, which is composed
by idler mode and signal mode resulting from a parametric down conversion
amplifier \cite{3-1}, is a typical entangled state of continuous
variable\cite{3-2}. Recently, many groups have devoted to the research on
nonclassical properties of two-mode squeezed state\cite{4,5,6,7,8}. Lee
\cite{4} and Cave' group \cite{5} have studied the nonclassical photon
statistics of two-mode squeezed states, respectively.

Theoretically, two-mode squeezed vacuum state is constructed by the two-mode
squeezing operator $S_{2}=\exp \left[  \lambda \left(  a_{1}a_{2}-a_{1}%
^{\dagger}a_{2}^{\dagger}\right)  \right]  $ acting on the vacuum state
$\left \vert 00\right \rangle $,
\begin{equation}
S_{2}\left \vert 00\right \rangle =\sec \text{h}\lambda \exp \left(  -a_{1}%
^{\dagger}a_{2}^{\dagger}\tanh \lambda \right)  \left \vert 00\right \rangle ,
\label{01}%
\end{equation}
where $\lambda$ is a squeezing parameter. Considering the coordinate and
momentum operators, respectively,
\begin{equation}
Q_{j}=\frac{1}{\sqrt{2}}\left(  a_{j}+a_{j}^{\dagger}\right)  ,P_{j}=\frac
{1}{i\sqrt{2}}\left(  a_{j}-a_{j}^{\dagger}\right)  , \label{02}%
\end{equation}
one can recast $S_{2}$ into the form
\begin{equation}
S_{2}=\exp \left[  i\lambda \left(  Q_{1}P_{2}+Q_{2}P_{1}\right)  \right]  .
\label{03}%
\end{equation}
In the state $S_{2}\left \vert 00\right \rangle $, the variances of two-mode
quadrature operators\cite{2}
\begin{equation}
X=\frac{1}{2}\left(  Q_{1}+Q_{2}\right)  ,Y=\frac{1}{2}\left(  Q_{1}%
+Q_{2}\right)  \label{04}%
\end{equation}
take the standard form, i.e.,
\begin{equation}
\left \langle 00\right \vert S_{2}^{-1}X^{2}S_{2}\left \vert 00\right \rangle
=\frac{e^{-2\lambda}}{4},\left \langle 00\right \vert S_{2}^{-1}Y^{2}%
S_{2}\left \vert 00\right \rangle =\frac{e^{2\lambda}}{4}. \label{05}%
\end{equation}

In the present paper, we shall extend the two-mode squeezed states to the
three-mode case and investigate their some nonclassical properties. Based on
the Fan's ideas \cite{9}, we construct the following unitary operator in the
three-mode Fock space%
\begin{equation}
S_{3}=\exp \left \{  i\lambda \left[  Q_{1}\left(  P_{2}+P_{3}\right)
+Q_{2}\left(  P_{1}+P_{3}\right)  +Q_{3}\left(  P_{1}+P_{2}\right)  \right]
\right \}  . \label{06}%
\end{equation}
Then a question naturally arises: Is the operator $S_{3}$ also a three-mode
squeezing operator? If yes, what is its corresponding squeezed state? what is
particular nonclassical properties of these squeezed states? Facing with these
questions, this work is arranged as follows. In Sec.2, to answer the above
questions, we first derive the normally ordered form of $S_{3}$ by virtue of
the technique of integration within an ordered product (IWOP) of operators and
analyze if the squeezing exists. Then the corresponding squeezed states
$S_{3}\left \vert \vec{\alpha}\right \rangle $ are obtained, where $\left \vert
\vec{\alpha}\right \rangle \equiv \left \vert \alpha_{1},\alpha_{2},\alpha
_{3}\right \rangle $ is a three-mode coherent state. In Secs.3 and 4, we
concentrate our attention on studying the higher-order squeezing and
higher-order sub-Possonian photon statistics in the state $S_{3}\left \vert
\vec{\alpha}\right \rangle $. We find that $S_{3}\left \vert \vec{\alpha
}\right \rangle $ can be squeezed to all even orders and the variances of the
three-mode quadrature operators exhibit stronger squeezing comparing with the
shown case in Ref.\cite{9}. So we call $S_{3}$ the three-mode enhancing
squeezing operator. We devote Sec.5 to evaluate the violation of
Bell-inequality for the state $S_{3}\left \vert \vec{\alpha}\right \rangle $
using the formalism of the Wigner representation in phase space based on the
parity operator and the displacement operation. Finally, a brief summary is
given in Sec.6.

\section{Normally ordered form of $S_{3}$}

In this section, we derive the normally ordered expansion $S_{3}$ and then
obtain the corresponding squeezed state. Because these operators $Q_{1}P_{2}$,
$Q_{1}P_{3}$, $Q_{2}P_{1}$, $Q_{2}P_{3}$, $Q_{3}P_{1}$, $Q_{3}P_{2}$ in
Eq.(\ref{06}) do not \ make up a closed Lie algebra, we utilize the IWOP
technique \cite{10} to disentangle $S_{3}$. For this purpose, we rewrite
$S_{3}$ using the matrix form as
\begin{equation}
S_{3}=\exp \left[  i\lambda \left(  Q_{1},Q_{2},Q_{3}\right)  \mathcal{A}\left(
P_{1},P_{2},P_{3}\right)  ^{T}\right]  =\exp \left[  i\lambda Q_{j}A_{jk}%
P_{k}\right]  \label{b1}%
\end{equation}
with
\begin{equation}
\mathcal{A}=\left(
\begin{array}
[c]{ccc}%
0 & 1 & 1\\
1 & 0 & 1\\
1 & 1 & 0
\end{array}
\right)  ,\text{ \ }j,k=1,2,3, \label{b2}%
\end{equation}
here and henceforth the repeated indices represent the Einstein summation notation.

Appealing to the Baker-Hausdorff formula%
\begin{equation}
e^{C}Be^{-C}=B+\left[  C,B\right]  +\frac{1}{2}\left[  C,\left[  C,B\right]
\right]  +\frac{1}{3}\left[  C,\left[  C,\left[  C,B\right]  \right]  \right]
+\cdots, \label{4}%
\end{equation}
where $C$ and $B$ are operators, one can calculate out
\begin{align}
S_{3}^{-1}Q_{k}S_{3}  &  =Q_{k}-\lambda Q_{j}\mathcal{A}_{jk}+\frac{1}%
{2}i\lambda^{2}\left[  Q_{i}\mathcal{A}_{ij}P_{j},Q_{l}\mathcal{A}%
_{lk}\right]  +\cdots \nonumber \\
&  =Q_{i}\left(  e^{-\mathcal{\lambda A}}\right)  _{ik}=\left(  e^{-\lambda
\mathcal{A}}\right)  _{ki}Q_{i}, \label{5}%
\end{align}
and
\begin{align}
S_{3}^{-1}P_{k}S_{3}  &  =P_{k}+\lambda Q_{i}\mathcal{A}_{kj}+\frac{1}%
{2}i\lambda^{2}\left[  P_{l}\mathcal{A}_{lk},Q_{i}\mathcal{A}_{ij}%
P_{j}\right]  +\cdots \nonumber \\
&  =\left(  e^{\lambda \mathcal{A}}\right)  _{ki}P_{i}. \label{6}%
\end{align}
It is implied from Eqs.(\ref{5}) and (\ref{6}) that the action of $S_{3}$ on
the three-mode coordinate eigenstate $\left \vert \vec{q}\right \rangle $%
,$\ $whose expression in Fock space is \cite{11}
\begin{equation}
\left \vert \vec{q}\right \rangle =\pi^{-3/4}\exp[-\frac{\vec{q}}{2}+\sqrt
{2}\widetilde{\vec{q}}a^{\dagger}-\frac{1}{2}\tilde{a}^{\dagger}a]\left \vert
\vec{0}\right \rangle , \label{10}%
\end{equation}
leads to
\begin{equation}
S_{3}\left \vert \vec{q}\right \rangle =\left \vert \Lambda \right \vert
^{1/2}\left \vert \Lambda \vec{q}\right \rangle ,\Lambda=e^{-\lambda \mathcal{A}%
},\left \vert \Lambda \right \vert \equiv \det \Lambda, \label{7}%
\end{equation}
where $\tilde{a}^{\dagger}=\left(  a_{1}^{\dagger},a_{2}^{\dagger}%
,a_{3}^{\dagger}\right)  $, $\widetilde{\vec{q}}=\left(  q_{1},q_{2}%
,q_{3}\right)  $ and tilde represents transposition of matrix. According to
properties of matrix $\mathcal{A}$ in Eq.(\ref{b2}), by expanding the
exponential term $e^{-\lambda \mathcal{A}}$, we get
\begin{equation}
\Lambda=e^{-\lambda \mathcal{A}}=\left(
\begin{array}
[c]{ccc}%
u_{1} & v_{1} & v_{1}\\
v_{1} & u_{1} & v_{1}\\
v_{1} & v_{1} & u_{1}%
\end{array}
\right)  , \label{8-0}%
\end{equation}
where%
\begin{equation}
u_{1}=\frac{1}{3}\left(  e^{-2\lambda}+2e^{\lambda}\right)  ,v_{1}=\frac{1}%
{3}\left(  e^{-2\lambda}-e^{\lambda}\right)  . \label{8}%
\end{equation}
Similarly,
\begin{equation}
\Gamma=e^{\lambda \mathcal{A}}=\left(
\begin{array}
[c]{ccc}%
u_{2} & v_{2} & v_{2}\\
v_{2} & u_{2} & v_{2}\\
v_{2} & v_{2} & u_{2}%
\end{array}
\right)  , \label{8-1}%
\end{equation}
where
\begin{equation}
u_{2}=\frac{1}{3}\left(  e^{2\lambda}+2e^{-\lambda}\right)  ,v_{2}=\frac{1}%
{3}\left(  e^{2\lambda}-e^{-\lambda}\right)  , \label{8-2}%
\end{equation}
which will be used in the rest of this article.

From Eq.(\ref{7}), it is found that $S_{3}$ has natural representation in the
coordinate basis $\left \vert \vec{q}\right \rangle $
\begin{equation}
S_{3}=\int d\vec{q}S_{3}\left \vert \vec{q}\right \rangle \left \langle \vec
{q}\right \vert =\left \vert \Lambda \right \vert ^{1/2}\int d\vec{q}\left \vert
\Lambda \vec{q}\right \rangle \left \langle \vec{q}\right \vert . \label{9}%
\end{equation}
It then follows that
\begin{equation}
S_{3}^{-1}Q_{k}S_{3}=\left \vert \Lambda \right \vert \int d\vec{q}\left \vert
\vec{q}\right \rangle \left \langle \Lambda \vec{q}\right \vert Q_{k}\int d\vec
{q}^{\prime}\left \vert \Lambda \vec{q}^{\prime}\right \rangle \left \langle
\vec{q}^{\prime}\right \vert =\left(  \Lambda Q\right)  _{k}%
\end{equation}
which is consistent with Eq.(\ref{5}). Substituting Eq.(\ref{10}) into
Eq.(\ref{9}) and using the IWOP technique as well as three-mode vacuum
projector $\left \vert \vec{0}\right \rangle \left \langle \vec{0}\right \vert
=\colon \exp \left[  -\tilde{a}^{\dagger}a\right]  \colon,$ we easily get the
normally ordered form of $S_{3}$ after integrating out of $\vec{q}$
\begin{align}
S_{3}  &  =\left(  \frac{\det \Lambda}{\det N}\right)  ^{1/2}\exp \left[
\frac{1}{2}\tilde{a}^{\dagger}\left(  \Lambda N^{-1}\tilde{\Lambda}-I\right)
a^{\dagger}\right] \nonumber \\
&  \times \colon \exp \left[  \tilde{a}^{\dagger}\left(  \Lambda N^{-1}-I\right)
a\right]  \colon \exp \left[  \frac{1}{2}\tilde{a}\left(  N^{-1}-I\right)
a\right]  , \label{11}%
\end{align}
where $N=\left(  1+\tilde{\Lambda}\Lambda \right)  /2$. Here we have used the
mathematical formula
\begin{equation}
\int d^{n}x\exp \left[  -\tilde{x}Wx+\tilde{x}v\right]  =\pi^{n/2}(\det
W)^{1/2}\exp \left(  \frac{1}{4}\tilde{v}W^{-1}v\right)  .
\end{equation}

Finally, applying the operator $S_{3}$ in Eq.(\ref{11}) on the three-mode
coherent state $\left \vert \vec{\alpha}\right \rangle \equiv \left \vert
\alpha_{1},\alpha_{2},\alpha_{3}\right \rangle $, we obtain the three-mode
squeezed coherent state, denoted by $S_{3}\left \vert \vec{\alpha}\right \rangle
$, as follows
\begin{equation}
S_{3}\left \vert \vec{\alpha}\right \rangle =\left(  \frac{\det \Lambda}{\det
N}\right)  ^{1/2}\exp \left[  \frac{1}{2}\tilde{a}^{\dagger}\left(  \Lambda
N^{-1}\tilde{\Lambda}-I\right)  a^{\dagger}+\tilde{a}^{\dagger}\left(  \Lambda
N^{-1}-I\right)  \alpha+\frac{1}{2}\tilde{\alpha}\left(  N^{-1}-I\right)
\alpha \right]  \left \vert \vec{\alpha}\right \rangle . \label{a1}%
\end{equation}
When $\alpha_{i}=0,i=1,2,3$, Eq.(\ref{a1}) reduces to the three-mode squeezed
vacuum state
\begin{equation}
S_{3}\left \vert \vec{0}\right \rangle =\left(  \frac{\det \Lambda}{\det
N}\right)  ^{1/2}\exp \left[  \frac{1}{2}\tilde{a}^{\dagger}\left(  \Lambda
N^{-1}\tilde{\Lambda}-I\right)  a^{\dagger}\right]  \left \vert \vec
{0}\right \rangle . \label{12}%
\end{equation}

\section{Higher-order squeezing property of $S_{3}\left \vert \vec{\alpha
}\right \rangle $}

It is well known that the concept of higher-order squeezing, introduced by
Hong and Mandel \cite{13}, is another aspect for revealing nonclassical
charactertics of a quantum state \cite{14}. When $2m$-th moment in a state is
less than that in the coherent state, this state is said to be squeezed to
order $2m$. Here, we evaluate the higher-order squeezed behavior of the
three-mode squeezed coherent state $S_{3}\left \vert \vec{\alpha}\right \rangle
$.

Firstly, we derive the expression of the $2m$-th moment of quadrature
operators in the state $S_{3}\left \vert \vec{\alpha}\right \rangle $. The
quadratures in the three-mode case are defined as
\begin{equation}
X_{3}=\frac{1}{\sqrt{6}}\sum_{j=1}^{3}Q_{j}=\frac{1}{2\sqrt{3}}\sum_{j=1}%
^{3}\left(  a_{j}+a_{j}^{\dagger}\right)  , \label{13-0}%
\end{equation}
and
\begin{equation}
Y_{3}=\frac{1}{\sqrt{6}}\sum_{j=1}^{3}P_{j}=\frac{1}{i2\sqrt{3}}\sum_{j=1}%
^{3}\left(  a_{j}-a_{j}^{\dagger}\right)  , \label{13-1}%
\end{equation}
obeying $\left[  X_{3},Y_{3}\right]  =\frac{1}{2}i$.

Letting $F=\sum_{j=1}^{n}\left(  \eta_{j}a_{j}+\kappa_{j}a_{j}^{\dagger
}\right)  $ and using $\left[  a_{j},a_{k}^{\dag}\right]  =\delta_{jk}$, one
can easily have
\begin{equation}
\exp \left(  \gamma \Delta F\right)  =\colon \exp \left(  \gamma \Delta F\right)
\colon \exp \left[  \frac{1}{2}\gamma^{2}\left(  \sum_{j=1}^{n}\eta_{j}%
\kappa_{j}\right)  \right]  , \label{b3}%
\end{equation}
where $\Delta F\equiv F-\left \langle F\right \rangle $, $\eta_{j}$, $\kappa
_{j}$ and $\gamma$ are $C$ number and $\colon \colon$denotes normal ordering.
By expanding both sides of Eq.(\ref{b3}) as power series in $\gamma$, and
equating coefficients of $(2m)!\xi^{2m}$, it is obtained that
\begin{equation}
\left(  \Delta F\right)  ^{2m}=\sum_{k=0}^{m}\frac{2m!}{\left(  2m-2k\right)
!k!}\left(  \frac{\sum_{j=1}^{n}\eta_{j}\kappa_{j}}{2}\right)  ^{k}%
\colon \left(  \Delta F\right)  ^{2m-2k}\colon, \label{b3-1}%
\end{equation}
which is a basic formula in calculating $2m$-th moment of two-mode fields.
Considering Eqs.(\ref{5}) and (\ref{6}), we calculate the $2m$-th moment of
the three-mode quadrature $X_{3}$ in the state $S_{3}\left \vert \vec{\alpha
}\right \rangle $, i.e.,
\begin{align}
\left \langle \Delta X_{3}\right \rangle ^{2m}  &  =\left \langle \vec{\alpha
}\right \vert S_{3}^{-1}\left(  X_{3}^{2}-\left \langle X_{3}\right \rangle
^{2}\right)  ^{m}S_{3}\left \vert \vec{\alpha}\right \rangle \nonumber \\
&  =\left(  \frac{1}{12}\right)  ^{m}\left \langle \vec{\alpha}\right \vert
\left \{  \Delta \sum_{k=1}^{3}\left[  \left(  e^{-\lambda \mathcal{A}}\right)
_{jk}\left(  a_{j}+a_{j}^{\dagger}\right)  \right]  \right \}  ^{2m}\left \vert
\vec{\alpha}\right \rangle \nonumber \\
&  =\left(  \frac{1}{12}\right)  ^{m}\left \langle \vec{\alpha}\right \vert
(\Delta Q)^{2m}\left \vert \vec{\alpha}\right \rangle , \label{15}%
\end{align}
\newline where
\begin{equation}
Q\equiv \sum_{k=1}^{3}\left[  \left(  e^{-\lambda \mathcal{A}}\right)
_{jk}\left(  a_{j}+a_{j}^{\dagger}\right)  \right]  . \label{15-1}%
\end{equation}

As a result of Eqs.(\ref{b3-1}) and (\ref{8-0}), we obtain
\begin{equation}
\left(  \Delta Q\right)  ^{2m}=\sum_{k=0}^{m}\frac{2m!}{\left(  2m-2k\right)
!k!}\colon \left(  \Delta Q\right)  ^{2m-2k}\colon \left(  \frac{\sum
_{i,j=1}^{3}\left(  e^{-2\lambda \mathcal{A}}\right)  _{ij}}{2}\right)  ^{k}.
\label{18}%
\end{equation}
Further, we have
\begin{equation}
\left \langle \vec{\alpha}\right \vert \colon \left(  \Delta Q\right)
^{2m-2k}\colon \left \vert \vec{\alpha}\right \rangle =\left \{  \left[
\sum_{k=1}^{3}\left(  e^{-\lambda \mathcal{A}}\right)  _{jk}\left(  \alpha
_{j}+\alpha_{j}^{\ast}\right)  \right]  ^{2}-\left \langle Q\right \rangle
^{2}\right \}  ^{m-k}. \label{18-1}%
\end{equation}
On the other hand,
\begin{equation}
\left \langle Q\right \rangle ^{2}=\left(  \left \langle \vec{\alpha}\right \vert
\colon \left \{  \sum_{k=1}^{3}\left[  \left(  e^{-\lambda \mathcal{A}}\right)
_{jk}\left(  a_{j}+a_{j}^{\dagger}\right)  \right]  \right \}  \colon \left \vert
\vec{\alpha}\right \rangle \right)  ^{2}=\left[  \sum_{k=1}^{3}\left(
e^{-\lambda \mathcal{A}}\right)  _{jk}\left(  \alpha_{j}+\alpha_{j}^{\ast
}\right)  \right]  ^{2}. \label{18-2}%
\end{equation}
By putting Eq.(\ref{18-2}) into Eq.(\ref{18-1}), it is easily obtained that
\begin{equation}
\left \langle \vec{\alpha}\right \vert \colon \left(  \Delta Q\right)
^{2m-2k}\colon \left \vert \vec{\alpha}\right \rangle =\delta_{mk}. \label{18-3}%
\end{equation}
Thus, Considering Eqs.(\ref{18}) and (\ref{18-3}), the final expression for
$2m$-order squeezing of $X_{3}$ in Eq.(\ref{15}) is
\begin{equation}
\left \langle \Delta X_{3}\right \rangle ^{2m}=\left(  \frac{1}{4}\right)
^{m}\left(  2m-1\right)  !!e^{-4m\lambda}, \label{c1}%
\end{equation}
where $\left(  2m-1\right)  !!=1\cdot3\cdot5\cdots(2m-1)$ and we have used the
result of Eq.(\ref{8-0}), i.e.,
\begin{equation}
\sum_{j,k=1}^{3}\left(  e^{-2\lambda \mathcal{A}}\right)  _{jk}=3e^{-4\lambda}.
\label{19}%
\end{equation}
Similarly, the $2m$-order squeezing for $Y_{3}$ is expressed as%
\begin{equation}
\left \langle \Delta Y_{3}\right \rangle ^{2m}=\left(  \frac{1}{4}\right)
^{m}\left(  2m-1\right)  !!e^{4m\lambda}. \label{24}%
\end{equation}
From Eqs.(\ref{c1}) and (\ref{24}), it is easily seen that when $\lambda=0$,
\begin{equation}
\left \langle \Delta X_{3}\right \rangle ^{2m}=\left \langle \Delta
Y_{3}\right \rangle ^{2m}=\left(  \frac{1}{4}\right)  ^{m}\left(  2m-1\right)
!! \label{b7}%
\end{equation}
which is just $2m$-th moment in the coherent state. According to the
definition of higher-order squeezing, $S_{3}\left \vert \vec{\alpha
}\right \rangle $ can be squeezed to all even orders.

In particular, when $m=1$, Eqs.(\ref{c1}) and (\ref{24}) reduce to the usual
quadrature squeezing for the state $S_{3}\left \vert \vec{\alpha}\right \rangle
$, namely,
\begin{equation}
\left \langle \Delta X_{3}\right \rangle ^{2}=\frac{1}{4}e^{-4\lambda
},\left \langle \Delta Y_{3}\right \rangle ^{2}=\frac{1}{4}e^{4\lambda},
\label{b9}%
\end{equation}
and
\begin{equation}
\left \langle \Delta X_{3}\right \rangle ^{2}\left \langle \Delta Y_{3}%
\right \rangle ^{2}=\frac{1}{16}, \label{b10}%
\end{equation}
which shows that $S_{3}$ is also a three-mode squeezing operator. In addition,
Eq.(\ref{b9}) clearly indicates that the squeezed states $S_{3}\left \vert
\alpha \right \rangle $ exhibit stronger squeezing $e^{-4\lambda}$ in one
quadrature than that $e^{-2\lambda}$ of the usual squeezed state in
Eq.(\ref{05}). This is a way of enhanced squeezing. That is the reason why we
call $S_{3}$ three-mode enhanced squeezing operator.

\section{Higher-order sub-Poissonian photon statistics in $S_{3}\left \vert
\vec{\alpha}\right \rangle $}

In this section, we focus on investigating higher-order sub-Poissonian
statistics in the state $S_{3}\left \vert \vec{\alpha}\right \rangle $. The
concept of higher-order sub-Poissonian statistics was introduced in
\cite{15,16,17} in terms of factorial moment.

By using $\left \langle R^{\left(  k\right)  }\right \rangle =\left \langle
R\left(  R-1\right)  \cdots \left(  R-k+1\right)  \right \rangle =\left \langle
a^{\dagger k}a^{k}\right \rangle $, with $R=a^{\dagger}a$, a parameter $P_{k}$
can be defined as
\begin{equation}
P_{k}=\frac{\left \langle a^{\dagger k}a^{k}\right \rangle }{\left \langle
a^{\dagger}a\right \rangle ^{k}}-1 \label{25}%
\end{equation}
with $k$ is a positive integer. For all $k\geq2$, $P_{k}=0$ corresponds to the
Poisson distribution. The negative or positive values of parameter $P_{k}$ for
$k>2$ indicate higher-order sub-Poissonian or higher-order super-Poissonian
statistics, respectively.

For the three-mode case of higher-order sub-Poissonian statistics in
$S_{3}\left \vert \vec{\alpha}\right \rangle $, analogue to the single-mode
case, we can define the joint operator $A=\frac{1}{\sqrt{3}}\sum_{j=1}%
^{3}a_{j}$ to study the joint sub-Poissonian distribution of $S_{3}\left \vert
\vec{\alpha}\right \rangle $. Referring to Eqs.(\ref{5}) and (\ref{6}), the
similar transformation of $A$, $A^{\dagger}$ under the operator $S_{3}$ is
\begin{equation}
S_{3}^{-1}AS_{3}=\frac{1}{\sqrt{2}}\left(  vA+uA^{\dagger}\right)  ,S_{3}%
^{-1}A^{\dagger}S_{3}=\frac{1}{\sqrt{2}}\left(  vA^{\dagger}+uA\right)  ,
\label{26}%
\end{equation}
where
\begin{equation}
u=e^{-2\lambda}+e^{2\lambda},v=e^{-2\lambda}-e^{2\lambda}. \label{26-1}%
\end{equation}

According to coherent completeness $\int \frac{d^{2}\vec{\beta}}{\pi^{3}%
}\left \vert \vec{\beta}\right \rangle \left \langle \vec{\beta}\right \vert =1$
with $\left \vert \vec{\beta}\right \rangle =\left \vert \beta_{1},\beta
_{2},\beta_{3}\right \rangle $, we acquire the following equation
\begin{align}
\left \langle A^{\dagger k}A^{k}\right \rangle  &  =\left \langle \vec{\alpha
}\right \vert \left(  S_{3}^{-1}A^{\dagger}S_{3}\right)  ^{k}\left(  S_{3}%
^{-1}AS_{3}\right)  ^{k}\left \vert \vec{\alpha}\right \rangle \nonumber \\
&  =\int \frac{d^{2}\vec{\beta}}{\pi^{3}}\left \langle \vec{\alpha}\right \vert
\left[  vA+uA^{\dagger}\right]  ^{k}\left \vert \vec{\beta}\right \rangle
\left \langle \vec{\beta}\right \vert \left[  uA+vA^{\dagger}\right]
^{k}\left \vert \vec{\alpha}\right \rangle . \label{27}%
\end{align}
It is easy to check $\left[  A,A^{\dagger}\right]  =1$, which is fit for the
following equation \cite{18} for $A$ in normal ordering form
\begin{equation}
\left(  uA+vA^{\dagger}\right)  ^{m}=\left(  -i\sqrt{\frac{uv}{2}}\right)
^{m}\colon H_{m}\left(  i\sqrt{\frac{u}{2v}}A+i\sqrt{\frac{v}{2u}}A^{\dagger
}\right)  \colon, \label{28}%
\end{equation}
where $H_{m}$ is the $m$-order Hermite polynomial. By putting Eq(\ref{28})
into \ref{27}), we have
\begin{align}
\left \langle A^{\dagger k}A^{k}\right \rangle  &  =\left(  \frac{-uv}%
{2}\right)  ^{k}\int \frac{d^{2}\vec{\beta}}{\pi^{3}}\left \langle \vec{\alpha
}\right \vert H_{k}\left(  i\sqrt{\frac{v}{2u}}A+i\sqrt{\frac{u}{2v}}%
A^{\dagger}\right)  \left \vert \vec{\beta}\right \rangle \left \langle
\vec{\beta}\right \vert H_{k}\left(  i\sqrt{\frac{u}{2v}}A+i\sqrt{\frac{v}{2u}%
}A^{\dagger}\right)  \left \vert \vec{\alpha}\right \rangle \nonumber \\
&  =\left(  \frac{-uv}{2}\right)  ^{k}\int \frac{d^{2}\vec{\beta}}{\pi^{3}%
}H_{k}\left(  \xi \right)  H_{k}\left(  \varsigma \right)  \exp \left[
-\sum_{j=1}^{3}\left(  \left \vert \beta_{j}\right \vert ^{2}+\left \vert
\alpha_{j}\right \vert ^{2}+\beta_{j}^{\ast}\alpha_{j}+\beta_{j}\alpha
_{j}^{\ast}\right)  \right]  \label{29}%
\end{align}
where
\begin{equation}
\xi \equiv i\left(  \sqrt{\frac{v}{6u}}\sum_{j=1}^{3}\beta_{j}+\sqrt{\frac
{u}{6v}}\sum_{j=1}^{3}\alpha_{j}^{\dagger}\right)  , \label{30}%
\end{equation}
and%
\begin{equation}
\varsigma \equiv i\left(  \sqrt{\frac{u}{6v}}\sum_{j=1}^{3}\alpha_{j}%
+\sqrt{\frac{v}{6u}}\sum_{j=1}^{3}\beta_{j}^{\ast}\right)  . \label{30-1}%
\end{equation}
Noticing that Hermite polynomial can be gotten from differential of mother
function, i.e.,
\begin{equation}
H_{m}\left(  x\right)  =\frac{\partial^{m}}{\partial t^{m}}\left.  \exp \left(
2xt-t^{2}\right)  \right \vert _{t=0}, \label{31}%
\end{equation}
Eq.(\ref{29}) can be simplified as
\begin{align}
\left \langle A^{\dagger k}A^{k}\right \rangle  &  =\frac{\left(  -uv\right)
^{k}}{2^{k}}\frac{\partial^{2k}}{\partial t_{1}^{k}\partial t_{2}^{k}}%
\int \frac{d^{2}\vec{\beta}}{\pi^{3}}\nonumber \\
&  \left.  \exp \left[  2\xi t_{1}-t_{1}^{2}+2\varsigma t_{2}-t_{2}^{2}%
-\sum_{j=1}^{3}\left(  \left \vert \beta_{j}\right \vert ^{2}+\left \vert
\alpha_{j}\right \vert ^{2}+\beta_{j}^{\ast}\alpha_{j}+\beta_{j}\alpha
_{j}^{\ast}\right)  \right]  \right \vert _{t_{1},t_{2}=0}. \label{32}%
\end{align}
Through integrating out variables $\vec{\beta},$ it is obtained that
\begin{equation}
\left \langle A^{\dagger k}A^{k}\right \rangle =\frac{\left(  -uv\right)  ^{k}%
}{2^{k}}\left.  \frac{\partial^{2k}}{\partial t_{1}^{k}\partial t_{2}^{k}}%
\exp \left(  -t_{1}^{2}-t_{2}^{2}+t_{1}G+t_{2}M-\frac{2v}{u}t_{1}t_{2}\right)
\right \vert _{t_{1},t_{2}=0}, \label{33}%
\end{equation}
where
\begin{equation}
G\equiv i\sum_{j=1}^{3}\left(  \sqrt{\frac{2u}{3v}}\alpha_{i}^{\ast}%
-\sqrt{\frac{2v}{3u}}\alpha_{j}\right)  ,M\equiv i\sum_{j=1}^{3}\left(
\sqrt{\frac{2u}{3v}}\alpha_{i}-\sqrt{\frac{2v}{3u}}\alpha_{j}^{\ast}\right)  .
\label{34}%
\end{equation}
A careful observation of Eq.(\ref{33}) reveals that if we get rid of
$-\frac{2v}{u}t_{1}t_{2}$ in the exponent, the rest just are the mother
function of Hermite polynomial. Fortunately we can express $\exp \left(
-\frac{2vt_{1}t_{2}}{u}\right)  $ in the following way
\begin{equation}
\exp \left(  -\frac{2vt_{1}t_{2}}{u}\right)  =\left.  \sum_{n=0}^{\infty}%
\frac{\left(  -2v\right)  ^{n}}{n!u^{n}}\frac{\partial^{2n}}{\partial
G^{n}\partial M^{n}}\exp \left[  -t_{1}^{2}-t_{2}^{2}+t_{1}G+t_{2}M\right]
\right \vert _{t_{1},t_{2}=0} \label{35}%
\end{equation}
Then substituting Eq.(\ref{35}) into Eq.(\ref{33}) and considering the
property of Hermite polynomial Eq.(\ref{31}), i.e.,
\begin{equation}
\frac{\partial}{\partial x}H_{n}\left(  x\right)  =2nH_{n-1}\left(  x\right)
. \label{36}%
\end{equation}
we get the expression for the average of $A^{\dagger k}A^{k}$ in the state
$S_{3}\left \vert \vec{\alpha}\right \rangle $
\begin{equation}
\left \langle A^{\dagger k}A^{k}\right \rangle =\frac{\left(  -uv\right)  ^{k}%
}{2^{k}}\sum_{n=0}^{k}\frac{\left(  -2v\right)  ^{n}\left(  k!\right)  ^{2}%
}{2^{4n}u^{n}\left[  \left(  k-n\right)  !\right]  ^{2}n!}H_{k-n}\left[
\frac{G}{2}\right]  H_{k-n}\left[  \frac{M}{2}\right]  . \label{37}%
\end{equation}
Especially, when $k=1$, noticing Eqs.(\ref{26-1}) and (\ref{37}), one has%
\begin{equation}
\left \langle A^{\dagger}A\right \rangle =\left[  GM-\frac{1}{8}\tanh \left(
-2\lambda \right)  \right]  \sinh \left(  4\lambda \right)  , \label{38}%
\end{equation}
where
\begin{equation}
GM=\frac{2}{3}\sum_{j,k=1}^{3}\left(  \alpha_{k}^{\ast}\alpha_{j}^{\ast
}+\alpha_{k}\alpha_{j}\right)  -\frac{4}{3}\coth \left(  -4\lambda \right)
\sum_{j,k=1}^{3}\alpha_{j}^{\ast}\alpha_{k}. \label{39}%
\end{equation}
For $k=2$,
\begin{align}
\left \langle A^{\dagger2}A^{2}\right \rangle  &  =\frac{\left(  uv\right)
^{2}}{2^{2}}\left[  \frac{v^{2}}{2^{5}u^{2}}-\frac{v}{2u}H_{1}\left(  \frac
{G}{2}\right)  H_{1}\left(  \frac{M}{2}\right)  +H_{2}\left(  \frac{G}%
{2}\right)  H_{2}\left(  \frac{M}{2}\right)  \right] \nonumber \\
&  =\sinh^{2}\left(  -4\lambda \right)  \left[  2^{-5}\tanh^{2}\left(
-2\lambda \right)  +2GM\tanh \left(  2\lambda \right)  +\left(  G^{2}-2\right)
\left(  M^{2}-2\right)  \right]  \label{40}%
\end{align}
where
\begin{equation}
G^{2}=\frac{4}{3}\sum_{j,k=1}^{3}\alpha_{k}^{\ast}\alpha_{j}-\frac{2}{3}%
\sum_{j,k=1}^{3}\left[  \alpha_{j}\alpha_{k}\tanh \left(  -2\lambda \right)
+\alpha_{j}^{\ast}\alpha_{k}^{\ast}\coth \left(  -2\lambda \right)  \right]  ,
\label{41}%
\end{equation}
and
\begin{equation}
M^{2}=\frac{4}{3}\sum_{j,k=1}^{3}\alpha_{k}^{\ast}\alpha_{j}-\frac{2}{3}%
\sum_{j,k=1}^{3}\left[  \alpha_{j}\alpha_{k}\coth \left(  -2\lambda \right)
+\alpha_{j}^{\ast}\alpha_{k}^{\ast}\tanh \left(  -2\lambda \right)  \right]  .
\label{41-1}%
\end{equation}

\begin{figure}[ptb]
\label{fig1} \centering
\includegraphics[width=10cm]{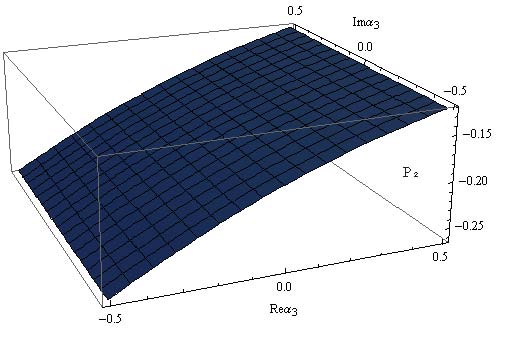}\caption{ Negative value of
$P_{2}$ changing with $\alpha_{3}$ by letting $\alpha
_{1}=\alpha_{2}=\lambda=1$ to reflect the sub-Poissonian
distribution of
$S_{3}\left \vert \vec{\alpha}\right \rangle $}%
\end{figure}Then introducing Eqs.(\ref{38}) and (\ref{40}) into (\ref{25}) for
$k=2,$ we easily obtain the result of $P_{2}$. In order to clearly obverse the
sub-Poissonian statistics in the state $S_{3}\left \vert \vec{\alpha
}\right \rangle ,$ we plot Fig.1 to visualize the change of $P_{2}$ with
$\alpha_{3}$ by simply letting $\alpha_{1}=\alpha_{2}=\lambda=1$. It clearly
shows that the parameter $P_{2}$ $<0$ when $-0.5<\operatorname{Re}\left(
\alpha_{3}\right)  <\,0.5$ but unchanges along with $\operatorname{Im}\left(
\alpha_{3}\right)  $. The negative value of $P_{2}$ displays the
sub-Poissonian distribution of $S_{3}\left \vert \vec{\alpha}\right \rangle $.
According to Ref.\cite{19}, the higher-order sub-Poissonian statistics may
become useful for their use in detection of higher-order squeezing.

\section{Bell-inequality Violation for $S_{3}\left \vert \vec{\alpha
}\right \rangle $}

The quantum nonlocality for continuous variable states has attracted much
attention. Wigner function representation of the Bell-inequality has been
developed using a parity operator as a quantum observable \cite{20,21,22}. In
this section, we turn our attention to evaluate the violation of the
Bell-inequality for squeezed coherent states $S_{3}\left \vert \vec{\alpha
}\right \rangle $ using the formalism of the Wigner representation in phase
space based on the parity operator and the displacement operation.

For the three-mode system, the correlation function is the expectation of the
operator
\begin{equation}
\Pi \left(  \vec{\beta}\right)  =%
{\displaystyle \bigotimes \limits_{j=1}^{3}}
D_{j}\left(  \beta_{j}\right)  \left(  -1\right)  ^{a_{j}^{\dagger}a_{j}}%
D_{j}^{\dagger}\left(  \beta_{j}\right)  \label{42}%
\end{equation}
which is a equivalent definition of the Wigner function, namely
\begin{equation}
W_{3}\left(  \vec{q},\vec{p}\right)  =W_{3}\left(  \vec{\beta}\right)
=\frac{1}{\pi^{3}}\left \langle \Pi \left(  \vec{\beta}\right)  \right \rangle
\label{43}%
\end{equation}
where $\vec{\beta}\equiv \left(  \beta_{1},\beta_{2},\beta_{3}\right)
=\frac{1}{\sqrt{2}}\left(  \vec{q}+i\vec{p}\right)  $ and $D_{j}\left(
\beta_{j}\right)  =\exp[\beta_{j}a_{j}^{\dagger}-\beta_{j}^{\ast}a_{j}]$ is
phase-space displacement operators acting on mode $j$. Thus $\Pi_{j}\left(
\beta_{j}\right)  $ is a product of displaced parity operators given as
\begin{equation}
\Pi_{j}\left(  \beta_{j}\right)  =D_{j}\left(  \beta_{j}\right)  \sum
_{n=0}^{\infty}\left(  \left \vert 2k\right \rangle \left \langle 2k\right \vert
-\left \vert 2k+1\right \rangle \left \langle 2k+1\right \vert \right)
D_{j}^{\dagger}\left(  \beta_{j}\right)  \label{44}%
\end{equation}
corresponding to the measurement of an even (parity $+1$) or an odd (parity
$-1$) number of photons in mode $j$. Within the framework of local realistic
theories, the Wigner representation of Bell inequality is of the form
\cite{23,24,25}
\begin{equation}
\left \vert B(n)\right \vert \leq2
\end{equation}
where $B(n)$ is combination of $W\left(  \vec{\beta}\right)  $. For the
three-mode case, $n$ $=3$, Bell-inequality can be expressed as
\begin{equation}
B(3)=\pi^{3}\left[  W(\beta_{1},\beta_{2},\beta_{3}^{\prime})+W(\beta
_{1},\beta_{2}^{\prime},\beta_{3})+W(\beta_{1}^{\prime},\beta_{2},\beta
_{3})-W(\beta_{1}^{\prime},\beta_{2}^{\prime},\beta_{3}^{\prime})\right]
\label{45}%
\end{equation}
Next, in order to examine the violation of the Bell-inequality for the state
$S_{3}\left \vert \vec{\alpha}\right \rangle $, we first derive the explicit
expression of its Wigner function. Recalling that in Refs.\cite{26,27,28} we
have introduced the Weyl ordering form of single-mode Wigner operator
$\Delta_{1}\left(  q_{1},p_{1}\right)  $
\begin{equation}
\Delta_{1}\left(  q_{1},p_{1}\right)  =%
\begin{array}
[c]{c}%
\colon \\
\colon
\end{array}
\delta \left(  q_{1}-Q_{1}\right)  \delta \left(  p_{1}-P_{1}\right)
\begin{array}
[c]{c}%
\colon \\
\colon
\end{array}
, \label{46}%
\end{equation}
and its normal ordering form is
\begin{equation}
\Delta_{1}\left(  q_{1},p_{1}\right)  =\frac{1}{\pi}\colon \exp \left[  -\left(
q_{1}-Q_{1}\right)  ^{2}-\left(  p_{1}-P_{1}\right)  ^{2}\right]  \colon.
\label{47}%
\end{equation}
where symbols $%
\begin{array}
[c]{c}%
\colon:\\
\colon:
\end{array}
$ denotes Weyl ordering. Note that the order of Bose operators $a_{1}%
,a_{1}^{\dagger}$ within a normally ordered product and a weyl ordered product
can be permuted. The Weyl ordering has a remarkable property, i.e., the
order-invariance of Weyl ordered operators under similar transformations which
means
\begin{equation}
\mathcal{U}^{\mathcal{-}1}{}%
\begin{array}
[c]{c}%
\colon \\
\colon
\end{array}
F\left(  a,a^{\dagger}\right)
\begin{array}
[c]{c}%
\colon \\
\colon
\end{array}
\mathcal{U=%
\begin{array}
[c]{c}%
\colon \\
\colon
\end{array}
}\left.  \mathcal{U}^{\mathcal{-}1}F\left(  a,a^{\dagger}\right)
\mathcal{U}\right.
\begin{array}
[c]{c}%
\colon \\
\colon
\end{array}
, \label{48}%
\end{equation}
as if the fence $%
\begin{array}
[c]{c}%
\colon:\\
\colon:
\end{array}
$ does not exist.

For three-mode case, the Weyl ordering form of the Wigner operator is
\begin{equation}
\Delta_{3}\left(  \vec{q},\vec{p}\right)  =\left.  \mathcal{%
\begin{array}
[c]{c}%
\colon \\
\colon
\end{array}
}\delta \left(  \vec{q}-\vec{Q}\right)  \delta \left(  \vec{p}-\vec{P}\right)
\mathcal{%
\begin{array}
[c]{c}%
\colon \\
\colon
\end{array}
}\right.  , \label{49}%
\end{equation}
where $\vec{Q}=\left(  Q_{1},Q_{2},Q_{3}\right)  $ and $\vec{P}=\left(
P_{1},P_{2},P_{3}\right)  $. Thus, according to the principle of Weyl ordering
invariance under similar transformations as well as Eqs.(\ref{5}), (\ref{6}),
(\ref{49}), we have
\begin{align}
W_{\vec{\alpha}}  &  =\left \langle \vec{\alpha}\right \vert S_{3}^{-1}%
\begin{array}
[c]{c}%
\colon \\
\colon
\end{array}
\delta \left(  \vec{q}-\vec{Q}\right)  \delta \left(  \vec{p}-\vec{P}\right)
\begin{array}
[c]{c}%
\colon \\
\colon
\end{array}
S_{3}\left \vert \vec{\alpha}\right \rangle \nonumber \\
&  =\left \langle \vec{\alpha}\right \vert \Delta \left(  e^{-\lambda \mathcal{A}%
}\vec{q},e^{\lambda \mathcal{A}}\vec{p}\right)  \left \vert \vec{\alpha
}\right \rangle \label{50}%
\end{align}
Further, Considering Eqs.(\ref{8-0}), (\ref{8-1}) and (\ref{47}), it is
obtained that
\begin{align}
W_{\vec{\alpha}}  &  =\frac{1}{\pi^{3}}\left \langle \vec{\alpha}\right \vert
\colon \exp \left[  -\left(  e^{-\lambda \mathcal{A}}\vec{q}-\vec{Q}\right)
^{2}-\left(  e^{\lambda \mathcal{A}}\vec{p}-\vec{P}\right)  ^{2}\right]
\colon \left \vert \vec{\alpha}\right \rangle \nonumber \\
&  =\frac{1}{\pi^{3}}\exp \left[  -\left(  e^{-\lambda \mathcal{A}}\vec{q}%
-\frac{1}{\sqrt{2}}\left(  \vec{\alpha}+\vec{\alpha}^{\ast}\right)  \right)
^{2}-\left(  e^{\lambda \mathcal{A}}\vec{q}-\frac{1}{i\sqrt{2}}\left(
\vec{\alpha}-\vec{\alpha}^{\ast}\right)  \right)  ^{2}\right] \nonumber \\
&  =\frac{1}{\pi^{3}}\exp \left[  -\Lambda_{ji}\Lambda_{ik}q_{j}q_{k}%
-\sum_{j=1}^{3}\sigma_{j}^{2}+2q_{k}\Lambda_{kj}\sigma_{j}-\Gamma_{ji}%
\Gamma_{ik}p_{j}p_{k}-\sum_{j=1}^{3}\chi_{j}^{2}+2p_{i}\Gamma_{kj}\chi
_{j}\right] \nonumber \\
&  =\frac{1}{\pi^{3}}\exp \left \{  -\sum_{j=1}^{3}\left[  \left(  u_{1}%
^{2}+2v_{1}^{2}\right)  q_{j}^{2}-\sigma_{j}^{2}+2u_{1}q_{j}\sigma_{j}\right]
-2\sum_{j>k=1}^{3}\left[  \left(  2u_{1}v_{1}+v_{1}^{2}\right)  q_{j}%
q_{k}+2v_{1}q_{j}\sigma_{k}\right]  \right. \nonumber \\
&  \left.  -\sum_{j=1}^{3}\left[  \left(  u_{2}^{2}+2v_{2}^{2}\right)
p_{j}^{2}-\chi_{j}^{2}+2u_{2}p_{j}\chi_{j}\right]  -2\sum_{j>k=1}^{3}\left[
\left(  2u_{2}v_{2}+v_{2}^{2}\right)  p_{j}p_{k}+2v_{2}p_{j}\chi_{k}\right]
\right \}  , \label{52}%
\end{align}
where $\sigma_{j}=\frac{1}{\sqrt{2}}\left(  \alpha_{j}+\alpha_{j}^{\ast
}\right)  $ and $\chi_{j}=\frac{1}{i\sqrt{2}}\left(  \alpha_{j}-\alpha
_{j}^{\ast}\right)  $. Especially, when $\alpha_{j}=0$, Eq.(\ref{52}) reduces
to the expression of Wigner function for the three-mode squeezed vacuum state
$S_{3}\left \vert \vec{0}\right \rangle $ in Eq.(\ref{12}).

\begin{figure}[ptb]
\label{fig2-1} \centering
\includegraphics[width=8cm]{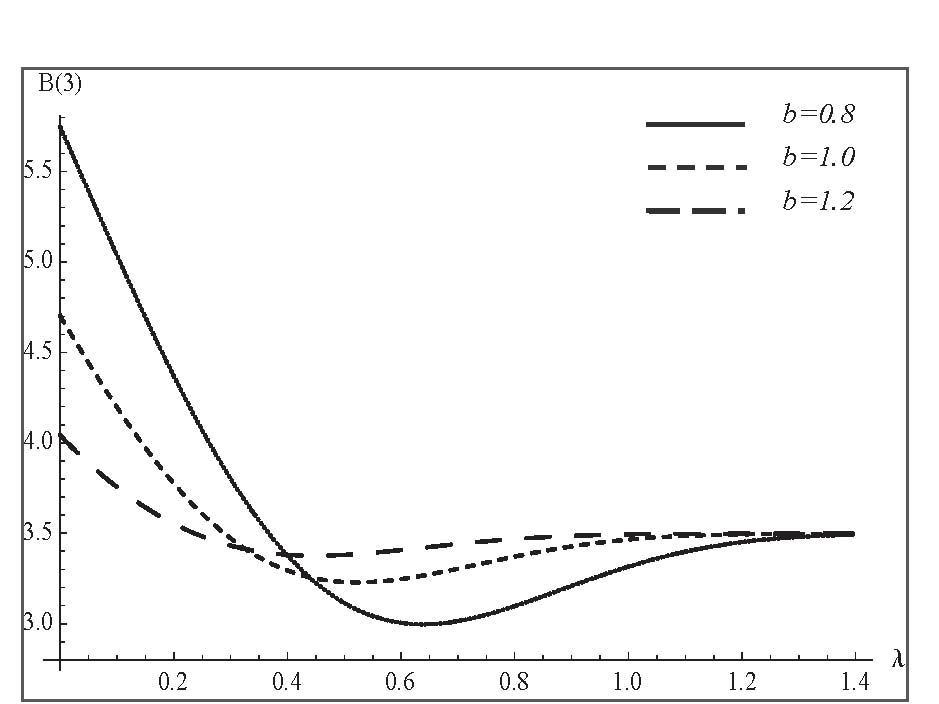}\caption{ Violation of the
Bell-inequality $B(3)$ for the state $S_{3}\left \vert
\vec{\alpha}\right \rangle $ by using parity measurements with $\alpha_{1}%
=0.4$, $\alpha_{2}=0.5$, $\alpha_{3}=0.6$, $\beta_{1}=\beta_{2}=\beta
_{3}^{\prime}=0$, $\beta_{3}=-b$, and $\beta_{1}^{\prime}=\beta_{2}^{\prime
}=b$. }%
\end{figure}

Now we put Eq.(\ref{52}) with corresponding variables into three-mode
Bell-inequality Eq.(\ref{45}). This Bell-inequality has $13$ variables and it
is highly nontrivial to find the global maximum values of $B\left(  3\right)
$ for all $13$ variables. To deal with this problem, in Fig.2 we plot the
maximal Bell violation of\ the state $S_{3}\left \vert \vec{\alpha
}\right \rangle $ as a function of $\lambda$ by defining $\alpha_{1}=0.4$,
$\alpha_{2}=0.5$, $\alpha_{3}=0.6$, $\beta_{1}=\beta_{2}=\beta_{3}^{\prime}%
=0$, $\beta_{3}=-b$, and $\beta_{1}^{\prime}=\beta_{2}^{\prime}=b$ ($b$ is a
positive constant associated with the displacement magnitude). From Fig.2, we
see that when the $\lambda \in(0,1)$, $B(3)>2$, the violation of the
Bell-inequality exists for positive squeezing parameter. Another interesting
phenomenon reflected by Fig.2 is that $B(3)$ will convergent to a constant
more than $2$ independent of change of $\lambda$.

\section{Conclusion}

In summary, we generalize the usual two-mode squeezed state to three-mode
operator $S_{3}$ and present the normal ordered form by the IWOP technique.
Then, we apply the operator $S_{3}$ on three-mode coherent state $\left \vert
\vec{\alpha}\right \rangle $ to get a new squeezed state $S_{3}\left \vert
\vec{\alpha}\right \rangle $. It is found that the state $S_{3}\left \vert
\vec{\alpha}\right \rangle $ displays enhanced higher-order squeezed property,
which proves that $S_{3}$ is an enhanced squeezing operator. We also formulate
the explicit expression of $P_{k}$ in order to describe higher-order
sub-Poissonian statistics by taking $k=2$ as an example. Finally, we discover
that the state $S_{3}\left \vert \vec{\alpha}\right \rangle $ violates the
Bell-inequality by using the formalism of the Wigner representation.

\section*{Acknowledgments}

This work was supported by the National Natural Science Foundation of China
under grant numbers 10775097 and 10874174.


\begin{thebibliography}{99}                                                                                               %


\bibitem {1}Dodonov, V.V. J. Opt. B: Quantum semiclass. Opt. \textbf{2002,} 4, R1-R33.

\bibitem {2}Loudon, R; Knight, P.L. J. Mod. Opt. \textbf{1987, }34, 709-759.

\bibitem {3}Caves, C.M. Phys. Rev. D \textbf{1981,} 23, 1693-1708.

\bibitem {3-0}Caves, C.M.; Schumaker, B.L. Phys. Rev. A \textbf{1985,} 31,
3068-3092; Phys. Rev. A \textbf{1985}, 31\textbf{,} 3093-3111.

\bibitem {3-1}Mandel, L.; Wolf, E. Optical Coherence and Quantum Optics;
Cambridge University Press, Cambridge, 1995

\bibitem {3-2}Hiroshima, T. Phys. Rev. A \textbf{2001,} 63, 022305-022308;
Milburn, G.J.; Braunstein, S.L. Phys. Rev. A \textbf{1999,} 60, 937-942; Ban,
M.: J. Opt. B: Quantum Semiclass. Opt. \textbf{1999,} 1\textbf{,} L9-L11.

\bibitem {4}Lee, C.T. Phys. Rev. A \textbf{1990,} 42, 1608-1616.

\bibitem {5}Caves, C.M.; Zhu, C.; Milburn, G.J.; Schleich, W. Phys. Rev. A
\textbf{1991}, 43, 3854-3861.

\bibitem {6}Selvadoray, M.; Kumar, M.S.; Simon, R. Phy. Rev. A \textbf{1994,}
49, 4957-4967.

\bibitem {7}Gantsog, T; Tana\^{s}, R. Phys. Lett. A \textbf{1991, }152, 251-256.

\bibitem {8}Selvadoray, M.; and Kumar, M.S. Opt. Commun. \textbf{1997,} 136, 125-134.

\bibitem {9}Fan, H.Y. and Yu, G.C. Phys. Rev. A \textbf{2002, }65, 033829-033836.

\bibitem {10}Fan, H.Y. J. Opt. B: Quantum Semiclass. Opt. \textbf{2003,} 5,
R147-R163; Fan, H.Y.; Lu, H.L.; Fan, Y. Ann. Phys. \textbf{2006,} 321, 480-494.

\bibitem {11}Fan, H.Y.; VanderLinde, J. Phys. Rev. A \textbf{1989}, 39, 1552-1555.

\bibitem {13}Hong, C.K.; Mandel, L. Phys. Rev. Lett. \textbf{1985, }54,
323-325; Phys. Rev. A \textbf{1985, }32, 974-982.

\bibitem {14}Marian, P. Phys. Rev. A \textbf{1991}, 44, 3325-3330; Duc, T.M.;
Noh, J. Opt. Comm. \textbf{2008}, 281, 2842-2848.

\bibitem {15}Lee, C.T. Phys. Rev. A\textbf{ 1990,} 41, 1721-1723.

\bibitem {16}Agarwal, G.S.; Tara, K. Phys. Rev. A \textbf{1992,} 46, 485-488.

\bibitem {17}Erenso, D.; Vyas, R.; Singh, S. J. Opt. Soc. Am. B \textbf{2002,
}19, 1471-1475.

\bibitem {18}Hu, L.Y.; Fan H.Y. J. Opt. Soc. Am. B \textbf{2008,} 25, 1955-1964.

\bibitem {19}Prakash, H.; Mishra, D.K. J. Phys. B: At. Mol. Opt. Phys.
\textbf{2006}, 39, 2291-2297.

\bibitem {20}Banaszek, K; W\'{o}dkiewicz, K. Phys. Rev. A \textbf{1998},
58\textbf{,} 4345-4347; Phys. Rev. Lett. \textbf{1999}, 82, 2009-2013.

\bibitem {21}Hyunseok, J; Nguyen, B.A. Phys. Rev. A \textbf{2006,} 74, 022104-022111.

\bibitem {22}Jacobsen, S.H; Jarvis, P.D. J. Phys. A: Math. Theor.
\textbf{2008,} 41, 365301-365316.

\bibitem {23}Van, L.P.; Braunstein, S.L. Phys. Rev. A \textbf{2001,} 63, 022106-022111.

\bibitem {24}Bell, J.S; Physics \textbf{1964,} 1, 195-200.

\bibitem {25}Bennett, C.H.; DiVincenzo, D.P.; Smolin, J.A.; Wootters, W.K.
Phys. Rev. A \textbf{1996, }54, 3824-3851.

\bibitem {26}Fan, H.Y. J. Phys. A \textbf{1992, }25, 3443-3447.

\bibitem {27}Fan, H.Y.; Fan, Y. Int. J. Mod. Phys. A \textbf{2002,} 17, 701-708.

\bibitem {28}Fan, H.Y. Mod. Phys. Lett. A \textbf{2000,} 15, 2297-2303.
\end{thebibliography}
\end{document}